\documentclass[12pt]{article}
\usepackage{graphicx}
\usepackage{color}
\usepackage{amsmath}
\usepackage{amssymb}
\newcommand{\be}{\begin{equation}}
\newcommand{\ee}{\end{equation}}
\newcommand{\bd}{\begin{displaystyle}}
\newcommand{\ed}{\end{displaystyle}}
\newcommand{\ba}{\begin{eqnarray}}
\newcommand{\ea}{\end{eqnarray}}
\DeclareMathSymbol{\lang}{\mathord}{symbols}{"68}
\DeclareMathSymbol{\rang}{\mathord}{symbols}{"69}
\DeclareMathSymbol{\openbra}{\mathord}{symbols}{"68}
\DeclareMathSymbol{\closeket}{\mathord}{symbols}{"69}
\DeclareMathOperator{\Rre}{Re}
\DeclareMathOperator{\Iim}{Im}

\newcommand{\ket}[1]{{| #1 \closeket}}

\newcommand{\aver}[1]{{\lang #1 \rang}}
\title{Nonlinear coherent dynamics of an atom in an optical lattice}
\author{V.Yu. Argonov and S.V. Prants\\ 
Laboratory of Nonlinear Dynamical Systems,\\
V.I.Il'ichev Pacific Oceanological Institute\\
of the Russian Academy of Sciences, 43 Baltiiskaya st.,\\
690041 Vladivostok, Russia}
\begin{document}
\maketitle
\begin{abstract}
We consider a simple model of lossless interaction between
a two-level single atom and a standing-wave single-mode laser field
which creates a one-dimensional optical lattice. Internal dynamics of
the atom is governed by the laser field which is treated to be classical
with a large number of photons. Center-of-mass classical atomic motion is
governed by
the optical potential and the internal atomic degree of freedom.
The resulting Hamilton-Schr\"odinger equations of motion are a
five-dimensional
nonlinear dynamical system with two integrals of motion, the total atomic
energy
and the length of the Bloch vector are conserved during the interaction.
In our previous papers the motion of the atom has been shown to be regular or
chaotic (in the sense of exponential sensitivity to small variations in
initial conditions and/or the system's control parameters) in dependence on
values of the control parameters, the atom-field detuning and recoil
frequency. At exact atom-field resonance, exact solutions for both the
external and internal atomic degrees of freedom can be derived.
The center-of-mass motion does not depend in this case on the internal
variables,
whereas the Rabi oscillations of the atomic inversion
 is a frequency modulated signal with the
frequency to be defined by the atomic position in the optical lattice.
We study analytically correlations between the Rabi oscillations  and
the center-of-mass motion in two limiting cases of a regular motion out off the 
resonance:
(1)  far-detuned atoms and (2)  fastly moving atoms. The main focus of the
paper is chaotic atomic motion that may be quantified strictly by positive
values of the maximal Lyapunov exponent. It is shown that atom, depending
on the value of its total energy, can either oscillate chaotically in a well
of the optical potential or fly ballistically with weak chaotic
oscillations of its momentum or wander in the optical lattice changing
the direction of motion in a chaotic way. In the regime of chaotic wandering
atomic motion is shown to have fractal properties. We find a useful tool
to visualize complicated atomic motion~--- Poincar\'e mapping of atomic
trajectories in an effective three-dimensional phase space onto planes
of atomic internal variables and momentum. The Poincar\'e mappings
are constructing using a  translational invariance of the standing laser
wave.
We find common features with typical non-hyperbolic
Hamiltonian systems~--- chains of resonant islands
of different sizes imbedded in a stochastic sea, stochastic layers,
bifurcations,
and so on. The phenomenon of sticking of atomic trajectories to boundaries
of regular islands, that should have a great influence to atomic transport
in optical lattices, is found and demonstrated numerically.

PACS 42.50.Vk, 05.45.Mt, 05.45.Xt
\end{abstract}

\section{Introduction}

Light exerts mechanical forces on matter. This hypothesis was suggested
by Kepler \cite{Kepler} in 1619 to explain a deviation of the comet's tails
flying nearby the Sun. It was Maxwell who in 1873 estimated the light
pressure, using his theory of electromagnetism \cite{Max}, and has shown that
it is very small. Peter Lebedev was the first who in 1899 measured the light
pressure on a macroscopic body \cite{Lebedev}. The first experiments on deviation
of microscopic particles by light have been carried out by W. Gerlach and
O. Stern \cite{GS22}, by P. Kapitza and P. Dirac \cite{KD33}, and by O. Frisch
\cite{F33}.

Manipulation of atomic motion with the help of laser beams, creating
an optical lattice, is one of the most fastly growing field of modern
physics (for a review see, for example, \cite{RMP}). There are different
theoretical and experimental aspects of this interaction including cooling
and trapping of atoms, Bose-Einstein condensation, quantum computing and
processing information with atoms.

In this paper we review our recent results on nonlinear coherent dynamics of a
single two-level atom in an optical lattice created in a one-dimensional cavity 
by two counterpropagating
laser waves. We are working in the strong-coupling regime and neglect all 
the losses. We show that even in a one-dimensional approximation the
atomic motion can be very complicated. We analyze both regular
and chaotic motion of atoms in a stationary standing-wave laser field
containing a large number of photons. It should be stressed that there
is a difference between various types of erratic atomic motion in an
optical lattice. {\it Chaotic motion} is strictly defined as a motion
of a deterministic nonlinear dynamical system that is exponentially sensitive
to small variations in the system's initial conditions or/and its control
parameters. There are different types of chaotic motion of atoms in an
optical lattice, including chaotic nonlinear oscillations of atomic center of
mass in a well of the optical potential, chaotic ballistic motion, when
the atomic
momentum oscillates chaotically around a value of the
average momentum, and the last but not least, {\it chaotic wandering}
of an atom  when it changes its direction of motion in a chaotic way
\cite{PK01,PS01,P02,Pr02,JETP,PEZ}. All the types of
chaotic atomic motion are quantified by positive values of the maximal
Lyapunov exponent.

In an optical lattice chaotic motion in the strict sense of this
notation may occur when there is no any kind of noise,
including atomic spontaneous emission which is a random process.
The respective deterministic atomic equations of motion are approximate ones,
but they are fundamental since spontaneous emission may be considered as a
quantum noise. {\it Random walking} is a kind of motion that occurs with
ultracold atoms which are detuned far away from the carrier laser frequency, so
their
internal degrees of freedom can be eliminated adiabatically. Because the
values of
the momentum of ultracold atoms are compared with the value of
the photon momentum,
each time after emitting a spontaneous photon atom gets a kick in a random
direction. This effect is a quantum analogue  of the classical random
walking.

In a general situation we should take into account both
the internal atomic motion and spontaneous emission events. In this
case, however, the equations of motion cease to be a deterministic dynamical
system because they include random terms and one may expect much more
complicated type of atomic motion which, besides of chaotic motion,
caused by the fundamental atom-field interaction, includes a purely
stochastic
component caused by random events of spontaneous emission. We have shown
recently
that in a range of the control parameters (detuning, laser intensity, and
recoil frequency) and initial conditions atoms may change their direction
of motion
erratically even if their momenta are much larger than  the photon momentum.
We will call this type of motion as {\it chaotic walking}.

The main aim of this paper is to describe different aspects of deterministic
atomic motion in an optical lattice, both regular and chaotic ones.
The effects of spontaneous emission on the atomic motion will be
considered in
a forthcoming paper.

\section{Hamilton-Schr\"odinger equations of motion}

We consider a two-level atom with mass $m_a$ and transition
frequency $\omega_a$, moving with the momentum $P$ along the axis $X$
in an ideal cavity through the standing laser wave with the field frequency $\omega_f$ 
and the wave vector $k_f$. In the frame,
rotating with the frequency $\omega_f$, the standard cavity QED Hamiltonian is
the following:
\begin{equation}
\hat H=\frac{\hat P^2}{2m_a}+\frac{1}{2}\hbar(\omega_a-\omega_f)\hat\sigma_z-
\hbar \Omega\left(\hat\sigma_-+\hat\sigma_+\right)\cos{k_f\hat X}.
\label{Jaynes-Cum}
\end{equation}
Here $\hat\sigma_{\pm, z}$ are the Pauli operators which describe the transitions
between lower, $\ket{1}$, and upper, $\ket{2}$, states.
$\Omega$ is the Rabi frequency which is proportional to the square
root of the number of photons in the wave $\sqrt{n}$. 
The standing-wave field must be strong enough ($n\gg 1$),
so we can neglect a back reaction of atoms on it and
consider the field classically.
For electronic degree of freedom the simple wavefunction is
\begin{equation}
\ket{\Psi(t)}=a(t)\ket{2}+b(t)\ket{1},
\label{Psi}
\end{equation}
where $a$ and $b$ are the complex-valued probability amplitudes to find the
atom in the states $\ket{2}$ and $\ket{1}$, respectively.
Using the Hamiltonian (\ref{Jaynes-Cum}), we get the Schr\"odinger equation 
\begin{equation}
\begin{array}{l}
\begin{displaystyle}
i\frac{da}{dt}=\frac{\omega_a-\omega_f}{2}a-\Omega b\cos k_fX,\
\end{displaystyle}
\\
\\
\begin{displaystyle}
i\frac{db}{dt}=\frac{\omega_f-\omega_a}{2}b-\Omega a\cos k_fX,\ 
\end{displaystyle}
\end{array}
\label{sysa}
\end{equation}
where the atomic position $X$ is considered as a parameter. Let us introduce
instead of the complex-valued probability amplitudes $a$ and $b$ the 
following new real-valued variables:  
\begin{equation}
\begin{displaystyle}
u\equiv 2\Rre\left(ab^*\right),
\quad
v\equiv -2\Iim\left(ab^*\right),
\quad
z\equiv \left|a\right|^2-\left|b\right|^2,
\end{displaystyle}
\label{uvz_def}
\end{equation}
which are the quadratures of the atomic dipole moment ($u$ and $v$) 
and the atomic population inversion,  $z$. 

In the process of emitting and
absorbing photons, atoms not only change their internal electronic states
but their external translational states change as well due to the photon
recoil. If the atomic average momentum is large as compared to the photon momentum
$\hbar k_f$, one can describe the translational degree
of freedom classically satisfying to classical Hamilton equations of motion. 
The dynamics in the strong-coupling regime is now governed by the Hamilton-Schr\"odinger
equations 
\begin{equation}
\begin{aligned}
\dot x&=\omega_r p,
\\
\dot p&=- u\sin x,
\\
\dot u&=\Delta v,
\\
\dot v&=-\Delta u+2 z\cos x,
\\                                         
\dot z&=-2 v\cos x,
\end{aligned}
\label{mainsys}
\end{equation}
where $x\equiv k_f\aver{\hat X}$ and $p\equiv \aver{\hat P}/\hbar k_f$ are classical 
atomic center-of-mass position and momentum, respectively.
Dot denotes differentiation with respect to dimensionless time $\tau\equiv \Omega t$.
The normalized recoil frequency, $\omega_r\equiv\hbar k_f^2/m_a\Omega\ll 1$,
and the atom-field detuning,
$\Delta\equiv(\omega_f-\omega_a)/\Omega$, are the control parameters.
The system has two integrals of motion, namely the total energy
\begin{equation}
W\equiv\frac{\omega_r}{2}p^2+U,
\label{W}
\end{equation}
where
\begin{equation}
U\equiv-u\cos x-\frac{\Delta}{2}z,
\label{U}
\end{equation}
is the potential energy, and the Bloch vector
\begin{equation}
u^2+v^2+z^2=1.
\label{R}
\end{equation}
The conservation of the Bloch 
vector length immediately follows from 
Eqs. (\ref{uvz_def}).

\begin{figure}[htb]
\begin{center}
\includegraphics[width=0.8\textwidth,clip]{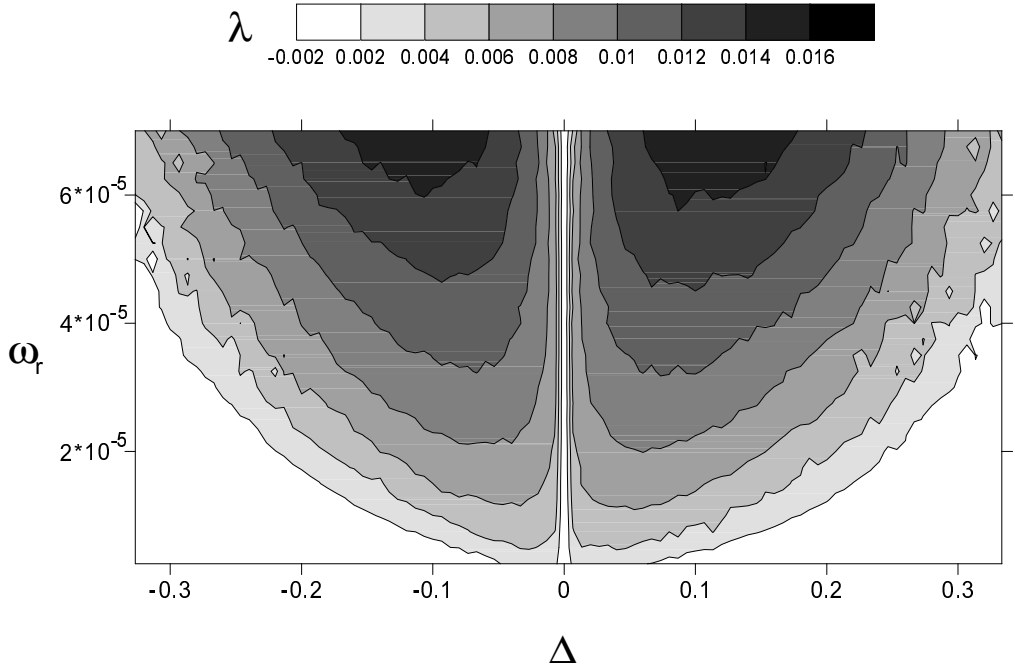}
\end{center}
\caption{Maximal Lyapunov exponent $\lambda$ vs the detuning $\Delta$
and the normalized recoil frequency $\omega_r$: $p_0=200$, $z_0=-1$, $u_0=v_0=0$.}
\label{fig1}
\end{figure}

Equations (\ref{mainsys}) with two integrals of motion constitute a Hamiltonian
autonomous system with two degrees of freedom and motion on a three-dimensional
hypersurface with a given energy value $W$. Generally, such a system has a positive 
Lyapunov exponent $\lambda$, a negative exponent equal in magnitude to positive one,
and two zero exponents. The sum of all Lyapunov exponents of a Hamiltonian system is
zero \cite{LL}. The maximal Lyapunov exponent characterizes the mean rate of the
exponential divergence of initially close trajectories,
\begin{equation}
\lambda=\lim\limits_{\tau\to\infty}\lambda(\tau),\quad
\lambda(\tau)=\lim\limits_{\delta(0)\to 0}\frac{1}{\tau}\ln\frac{\delta(\tau)}{\delta(0)},
\end{equation}
and serves as a quantitative measure of dynamical chaos in the system.
Here, $\delta(\tau)$ is a distance (in the Euclidian sense) at time
$\tau$ between two trajectories close to each other at initial time
$\tau=0$. The dependence of $\lambda$ on control parameters has been 
calculated in \cite{PK01, PS01} with the similar system. It has been              
shown that dynamical chaos in a strongly-coupled atom-field system
exist in a wide range of parameters and initial atomic momentum $p_0$.
The result of computation
of the maximal Lyapunov exponent with our system (\ref{mainsys}) 
in the space of
control parameters, $\omega_r$ and $\Delta$, is shown in
Fig.~\ref{fig1}.
In white regions of the map maximal
Lyapunov exponent $\lambda$ is almost zero and the dynamics is stable. In
other regions the positive values of $\lambda$ show the Lyapunov
instability. 

In all the numerical simulations we use physically realistic
parameters and initial conditions. For example, we can chose cesium atoms
with the transition wavelength $\lambda_a\simeq 852$~ nm. 
The Rabi frequency $\Omega$ depends on many factors including the field strength,
which could be changed in a wide range. 
In most computations we shall use the Rabi frequency of $\Omega\approx 10$~ GHz.
With this value of the Rabi frequency we get the normalized
recoil frequency to be $\omega_r=10^{-5}$. Also we put the initial
position $ x_0=0$. The detuning $\Delta$ could be varied
in a wide range, and the Bloch variables are restricted by the length
of the Bloch vector (\ref{R}). The most interesting effects are observed
with rather cold atoms. For example, $p_0=200$ taken in computing $\lambda$ 
in Fig.~\ref{fig1} with our normalization 
corresponds to the atomic velocity $v_a\approx 0.7$ m/s.
It should be noted that  we use in this paper the normalization to the laser 
Rabi frequency $\Omega$,  not to the vacuum (or single-photon) Rabi frequency 
as it has been done in our previous papers \cite{PK01,PS01,P02,Pr02,JETP}. 
So the ranges of the normalized control parameters, taken in this paper, 
differ from those in the cited papers.

\section{Regular dynamical regime}

\subsection{Exact atom-field resonance}

At exact resonance, $\Delta=0$, one can easily find an additional integral
of motion,
\be
u=const=u_0.
\end{equation}
In this case the fast and slow variables are separated from each other
allowing one to integrate exactly the reduced equations of motion.
Total energy becomes equal to
\be
W_R=\frac{\omega_r}{2}p^2-u_0\cos x,
\label{8}
\end{equation}
and the potential energy gets the simple form
\be
U_R=-u_0\cos x.
\label{UR}
\end{equation}
The center-of-mass translational motion of the atom in such a
spatially periodic potential of the standing wave
is described by the simple nonlinear equation for a free physical pendulum
\begin{equation}
\label{7}
\ddot x+\omega_r u_0\sin x=0,
\end{equation}
and does not depend on evolution of the internal degrees of freedom. 

\begin{figure}[ht]
\begin{center}
\includegraphics[width=0.9\textwidth,clip]{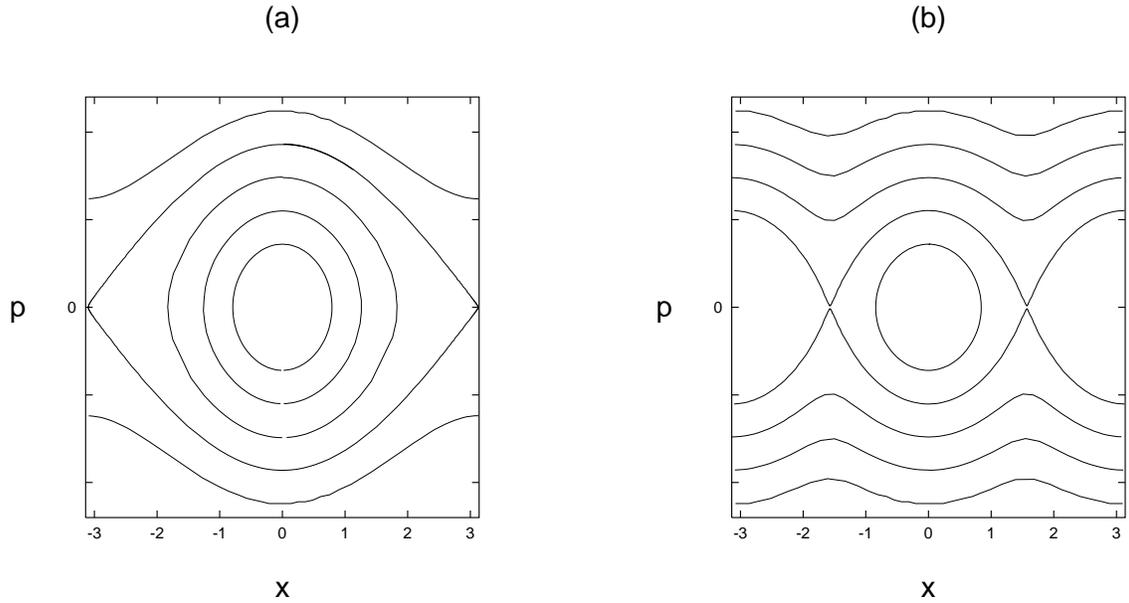}
\end{center}
\caption{ The typical regular phase portraits for the translational
degree of freedom: (a) $\Delta=0$; (b) $|\Delta|\gtrsim 0.2$.}
\label{fig2}
\end{figure}

The translational motion is trivial when $u_0$ is zero. In spite of the zero
potential field, a structure of a 
standing wave is still present in a cavity. In this case, the atom
will move in one direction with a constant velocity, and
the Rabi oscillations modulated by the standing wave will occur.
In general case one can easily get from (\ref{8}) the dependence $p( x)$
\be
p=\sqrt{\frac{2}{\omega_r}(W_R+u_0\cos x)},
\end{equation}
which gives the phase portrait of the system in the plane $( x, p)$
(Fig.~\ref{fig2}a). It is the phase portrait of a nonlinear pendulum
with three types of trajectories depending on the value of its energy
$W_R$: oscillator-like motion in a potential well if $W_R< u_0$,
a separatrix if $W_R= u_0$, and ballistic-like motion if $W_R> u_0$. 

For the initial values $ x_0=0$ and $\dot x_0=\omega_r p_0$,
the equation for the translational motion (\ref{7}) has the solution
\be
 x(\tau)=
\left\{
\begin{array}{ll}
\bd
2\arcsin\left[K\ \mbox{sn}\left[\sqrt{\omega_r u_0}\tau,
K\right]\right],\ed &\bd K^2\leqslant 1;\ed \\
\\
\bd
2\ \mbox{am}\left[\frac{1}{2}\omega_r p_0\tau,
\frac{1}{K}\right],\ed &\bd K^2\geqslant 1,\ed
\end{array}
\right.
\label{xi}
\end{equation}
\be
 p(\tau)=
\left\{
\begin{array}{ll}
\bd
 p_0\ \mbox{cn}\left[\sqrt{\omega_r u_0}
\tau, K\right],\ed &\bd K^2\leqslant 1;\ed\\
\\
\bd
 p_0\ \mbox{dn}\left[\frac{1}{2}\omega_r p_0
\tau,\frac{1}{K}\right],\ed &\bd K^2\geqslant 1,\ed
\end{array}
\right.
\end{equation}
where
\be
K=\frac{p_0}{2}\sqrt\frac{\omega_r}{u_0}
\label{11}
\end{equation}
is the modulus of the elliptic Jacobi functions. 
The solution gives the critical value of the atomic momentum
\be
p_{cr}=2\sqrt{u_0/\omega_r}.
\label{10}
\end{equation}
Atoms with $p_0\leqslant p_{cr}$ are trapped by the standing-wave
field, the result that is well-known from early studies \cite{Letokhov}.
The modulus $K$ is simply
connected with the normalized value of the difference between the
energy of the atom and its value on the separatrix
\be
K^2=1+\frac{W_R-u_0}{2u_0}.
\label{13}
\end{equation}

\begin{figure}[ht]
\begin{center}
\includegraphics[width=0.8\textwidth,clip]{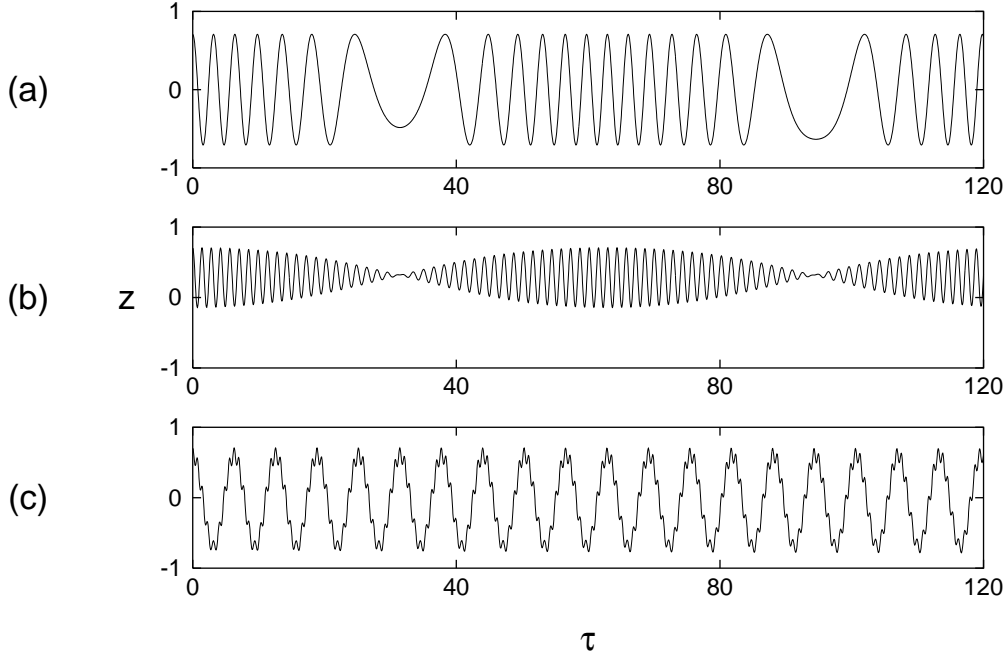}
\end{center}
\caption{Rabi oscillations: (a) frequency modulation at exact resonance,
$\Delta=0$, $p_0=5000$; (b) amplitude modulation far from resonance, $\Delta=-4$,
$p_0=5000$; (c) Doppler-Rabi resonance, $\Delta=-4$, $p_0=400000$.
In all panels, $z_0=u_0=\sqrt{0.5}$, $v_0=0$.}
\label{fig3}
\end{figure}

As to internal atomic evolution, it depends on the translational degree of freedom
since the force of the atom-field coupling depends on the position of the
atom in a periodic standing-wave potential. The equation for the atomic
population inversion $z(\tau)$ is derived from the two last 
equations of the set (\ref{mainsys}) with $\Delta=0$:
\be
\dot z=\mp 2\sqrt{1-z^2-u_0^2}\cos[ x(\tau)],
\label{15}
\end{equation}
where $\cos[ x(\tau)]$ is a known function of
the translational variables only which can be found with the help of 
the exact solutions obtained. It is easy to find the exact solution of
Eq. (\ref{15})
\begin{equation}
\label{16}
z(\tau)=\mp\sqrt{1-u_0^2}\ \sin\left(2\int_0^\tau \cos  x d\tau'+
\psi_0\right),
\end{equation}
where the sign is  opposite to that for the initial value $z_0$ and
\be
\psi_0=\mp\arcsin\frac{z_0}{\sqrt{1-u_0^2}} 
\end{equation}
is an integration constant. The internal energy
of the atom could be considered as a frequency-modulated signal
with the instant frequency $2\cos[ x(\tau)]$ and the modulation frequency
$\dot x=\omega_r p(\tau)$,
but it is correct only if the first value is much greater than the second,
i. e. $|\omega_r p_0|\ll 2$. Such a signal is shown in Fig.~\ref{fig3}a
for a ballistic atom ($p_0=5000$, $v_a\approx 17.5$ m/s).
With $|\omega_r p_0|\geqslant 2$
the modulation disappears and the signal becomes a periodic one with
the frequency $\omega_r p$. With fast atoms, $| \omega_r p_0 | \gg 2$
and $p\simeq p_0\gg p_{cr}$ (Raman-Nath approximation) Eq. (\ref{16}) is simplified
\begin{equation}
z(\tau)\approx z_0-\frac{2v_0}{\omega_r p_0}\sin\omega_r p_0\tau-\frac{4z_0}
{\omega_r^2 p_0^2}\sin^2\omega_r p_0\tau.
\label{zfast}
\end{equation}

\subsection{Non-resonant Rabi oscillations}

With comparatively small detunings $\Delta$ the dynamics of slow atoms
can be chaotic. In this case the Rabi oscillations are
still a signal with a frequency modulation, but
the amplitude is not constant anymore, it jumps chaotically
with the characteristic time $1/\omega_r p$.
With comparatively large detunings, the Rabi oscillations become regular,
but with a prominent periodic amplitude modulation, while 
the frequency modulation is not so deep (Fig.~\ref{fig3}b). 

In two limit cases, $|\Delta|\gg\mbox{max}[|\omega_r p|, 2]$
and $|\omega_r p|\gg\mbox{max}[|\Delta|, 2]$, the analytic solutions
can be obtained. In this both cases $u$ and $v$ 
are the harmonic functions with the frequency $\Delta$, and
for the internal energy we have 
\begin{equation}
z\approx
\left\{
\begin{array}{ll}\bd
z_0+\frac{2u_0}{\Delta}-\frac{2\sqrt{u_0^2+v_0^2}}{\Delta}\cos x\sin(\Delta\tau+\varphi_0),\ed&
|\Delta|\gg\mbox{max}[|\omega_r p|, 2],\\
\\
\bd z_0+
\frac{2\sqrt{u_0^2+v_0^2}}{\omega_r p_0}\cos(\Delta\tau+\varphi_0)\sin\omega_r p_0\tau,\ed&
|\omega_r p_0|\gg\mbox{max}[|\Delta|, 2],\ \omega_r p_0^2\gg 4,
\end{array}
\label{zz}
\right.
\end{equation}
where $\varphi_0=\arcsin(u_0/\sqrt{u_0^2+v_0^2})$.
The Rabi oscillations now are amplitude-modulated
signals with two characteristic frequencies $|\omega_r p|$ and $|\Delta|$.
The larger frequency is the main frequency and the other one is
the modulation frequency. In the solution for fast atoms
we also used the Raman-Nath approximation, $x\simeq\omega_r p_0\tau$, that
is correct if the initial kinetic energy $\omega_r p^2_0/2$ is much greater
than the amplitude of potential energy variations (equal to 2 in our case). 
With $\Delta=0$, the solution has the form 
(\ref{zfast}), but without the last term which is small.
Solutions (\ref{zz}) show a good correspondence with
the numerical experiments, performed in \cite{JETP} with
the similar equations. The typical amplitude-modulated Rabi oscillations
are shown in Fig.~\ref{fig3}b.

More exact solution for $u$ can be found
using the approximation $z\approx const\approx z_0$
(i. e. $z_{\rm max}-z_{\rm min}\ll|z_0|$), which is 
correct in the both limit cases considered above, excluding $z_0\approx 0$.
Then from Eq.~(\ref{mainsys}) we get the equation for a driven linear
oscillator
\begin{equation}
\ddot u+\Delta^2 u\approx 2z_0\Delta\cos x,
\label{ddotu}
\end{equation}
which has the solution
\begin{equation*}
u(\tau)\approx 2z_0\sin\Delta\tau\int\cos\Delta\tau\cos x d\tau
-2z_0\cos\Delta\tau\int\sin\Delta\tau\cos x d\tau+
\end{equation*}
\begin{equation}
+u_0\cos\Delta\tau+v_0\sin\Delta\tau.
\label{utau}
\end{equation}
For $|\Delta|\gg|\omega_r p|$, the solution (\ref{utau})
can be approximated as follows:
\begin{equation}
u\approx\frac{2z_0}{\Delta}\cos x+\sqrt{u_0^2+v_0^2}\ \sin(\Delta\tau+\varphi_0).
\label{utau2}
\end{equation}
Using Eqs.~(\ref{zz}) and (\ref{utau2}), we get the periodic
potential with the spatial period $\pi$ (in difference from
the resonant potential with the period $2\pi$):
\begin{equation}
U\approx-\frac{2z_0}{\Delta}\cos^2 x+const.
\label{Upi}
\end{equation}
The corresponding phase portrait is shown in Fig.~\ref{fig2}b.

When the frequencies are close, $|\omega_r p|\simeq|\Delta|$,
the Doppler-Rabi resonance takes place \cite{JETP} in spite of the fact
that the detuning may be very large.
Let us consider the standing wave as a combination
of two counter-propagating waves. In the frame, moving with
the atomic velocity, their frequencies, $\omega_1$ and $\omega_2$,
are different because of the Doppler effect:
\be
\omega_1=\omega_f-\frac{v_a}{c}\omega_f,\quad\quad\omega_2=\omega_f+\frac{v_a}{c}
\omega_f,
\end{equation}
where $v_a$ is the atomic velocity and $c$ is the speed of light.
The atom is rather slow so we can neglect the relativistic effects.
Let us consider atoms fast enough for the
Raman-Nath approximation $ p\approx p_0$ to be valid.
Renormalizing all the frequencies to $\Omega$,
we define the dimensionless detunings between the atomic transition
and the running wave frequencies as:
\be
\Delta_1\equiv\frac{\omega_1-\omega_a}{\Omega}=\Delta-\omega_r p_0,\quad\quad
\Delta_2\equiv\frac{\omega_2-\omega_a}{\Omega}=\Delta+\omega_r p_0.
\end{equation}
The condition $|\Delta|=|\omega_r p_0|$ leads
to the resonance between the atom and one of the waves.
If $|\Delta|\gg 1$, we can neglect the interaction with the other wave
and consider the atom as if only one wave with the frequency
$\omega_1$ or $\omega_2$ exists. In the field of the wave, say, $\omega_1$,
the dynamics can be described
by the Bloch-like equations
\begin{equation}
\begin{array}{c}
\dot u=\Delta_1 v,\quad \dot v=-\Delta_1 u+z,\quad \dot z=-v,
\end{array}
\label{34}
\end{equation}
in which the interaction energy does not depend on the atomic position
and its amplitude value is twice smaller as compared to the standing wave.
Eqs. (\ref{34}) have the solution
\be
z=\frac{u_0\Delta_1}{\omega_z^2}(1-\cos\omega_z\tau)-\frac{v_0}{\omega_z}
\sin\omega_z\tau+z_0\left( \frac{\Delta_1^2}{\omega_z^2}+\frac{1}{\omega_z^2}
\cos\omega_z\tau\right),
\label{zdr}
\end{equation}
where $\omega_z\equiv\sqrt{\Delta_1^2+1}=\sqrt{(\Delta-\omega_r p_0)^2+1}$.
At the exact Doppler resonance ($\Delta_1=0$), atomic internal
energy $z$ oscillates with the dimensionless frequency 
$1$, and the amplitude of oscillations is maximal.
Numerical simulations with Eqs.~(\ref{mainsys}) shows that
this speculations are correct (Fig.~\ref{fig2}c, where $p_0=400000$ and $v_a\approx 1400$ m/s),
and even very far from the resonance $\Delta=0$ the deep Rabi oscillations
can be observed for the atoms to be fast enough.

\section{Irregular dynamics: chaos and fractals}

\subsection{Chaotic atomic wandering}

\begin{figure}
\begin{center}
\includegraphics[width=0.8\textwidth,clip]{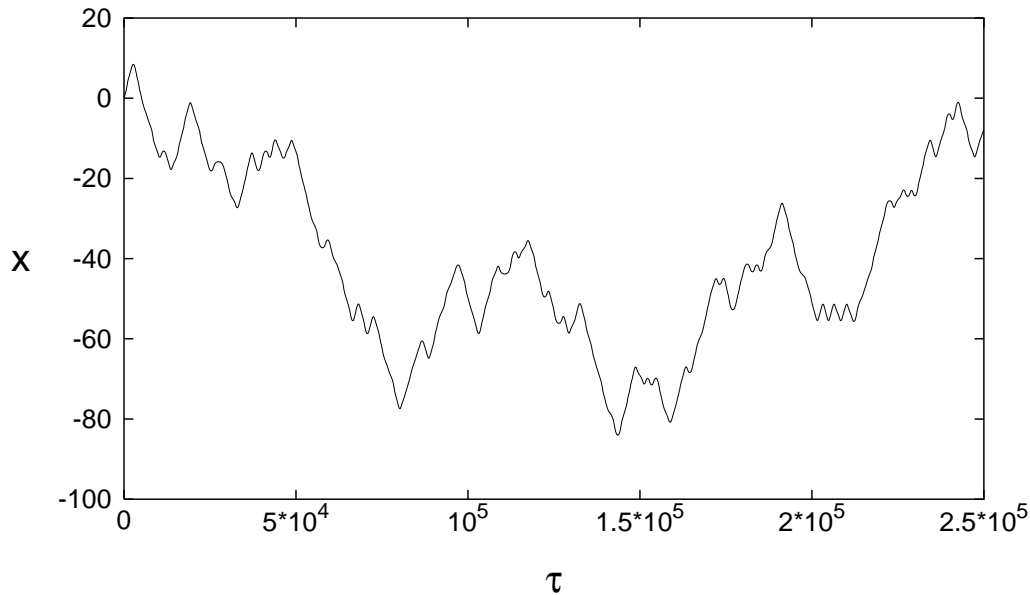}
\end{center}
\caption{A typical atomic trajectory in the regime of chaotic
wandering: $x_0=0$, $p_0=300$, $z_0=-1$, $u_0=v_0=0$, $\omega_r=10^{-5}$,
$\Delta=-0.05$.}
\label{fig4}
\end{figure}

In Fig.~\ref{fig1} we depict the maximal Lyapunov exponent map
in the space of control parameters, $\omega_r$ and $\Delta$.
The maximal Lyapunov exponent depends not only on
the parameters $\omega_r$ and $\Delta$, but on initial conditions
of the system (\ref{mainsys}), as well. Especially important is a value of
the initial momentum, $p_0$. The most interesting effects can be
observed with rather cold atoms, when the initial atomic kinetic
energy is close to the amplitude of the optical potential. In this case
we get the chaotic wandering of an atom in the standing wave. A typical
chaotic atomic trajectory is shown in Fig.~\ref{fig4}. 

It follows from (\ref{mainsys}) that the translational motion of the atom
at $\Delta \neq 0$ is described by the equation of a nonlinear physical
pendulum with a frequency modulation
\begin{equation}
\ddot x+\omega_r  u(\tau)\sin x=0,
\label{12}
\end{equation}
in which $u$ is the function of all the other dynamical variables.
The normalized Rabi oscillation frequency is a value of the order
of $\omega_z'\equiv\sqrt{\Delta^2+4}$ which substantially exceeds the frequency
of small-amplitude
translational motion $\sqrt{\omega_r  u_0}\ll 1$ in the potential well. Taking this into consideration,
the mechanism of the arising of chaos can be revealed \cite{P02}. The stochastic
layer width was estimated as
\begin{equation}
D\simeq 8\pi\left(\frac{\omega_z'}{\omega_0}\right)^3
\exp\left(\frac{-\pi\omega_z'}{2\omega_0}\right),
\end{equation}
where $\omega_0\equiv\sqrt{2\omega_r |\Delta|}/\omega_z'$, $\omega_z'/\omega_0\gg 1$.
The $D$ value is the energy change in the neighbourhood of the unperturbed
separatrix normalized with respect to the pendulum separatrix energy
$\omega_0^2$. Small changes in the energy causes comparatively small changes
in the frequency of oscillations. For the energies of motion that are strongly
different from the separatrix energy, that is, close to potential well
bottoms and high above optical potential $U$ hills, small frequency changes
cause small phase changes during the translation motion period. However, close to
the unperturbed separatrix, where the period of oscillations tends to infinity,
even small frequency changes can cause substantial phase changes. This is the
reason for the exponential instability of motion of the parametric nonlinear
oscillator (\ref{12}) and chaotic atomic motion in the field of a periodic
standing wave.

A clear idea of the character of chaotic wandering can be developed
using the model of ``two potentials''. At resonance,
the optical potential $U$ reproduces the structure of the standing wave in the cavity
(\ref{UR}) with the $2\pi$ period (the phase portrait in Fig.~\ref{fig2}a).
Far from the resonance, the potential has the period $\pi$ and is
approximately described by Eq. (\ref{Upi}) and the corresponding phase portrait
is shown in Fig.~\ref{fig2}b. These potentials will be called resonant and nonresonant,
respectively. We can say that, when the motion in the cavity is chaotic,
the both potentials ``virtually'' coexist. The well depths in both structures
change as time passes, and an atom randomly gets into one or another
structure every time when it crosses a standing wave node. The
probability of getting into the resonant or nonresonant potentials
depends on the detuning. Near the resonance atom is in the resonant potential
almost the whole time and only rarely gets into nonresonant one for a short time.

In our study
\cite{JETP} we have shown that chaotic wandering has fractal properties.

\subsection{Dynamical atomic fractals}

\begin{figure}[htb]
\begin{center}
\includegraphics[width=0.7\textwidth,clip]{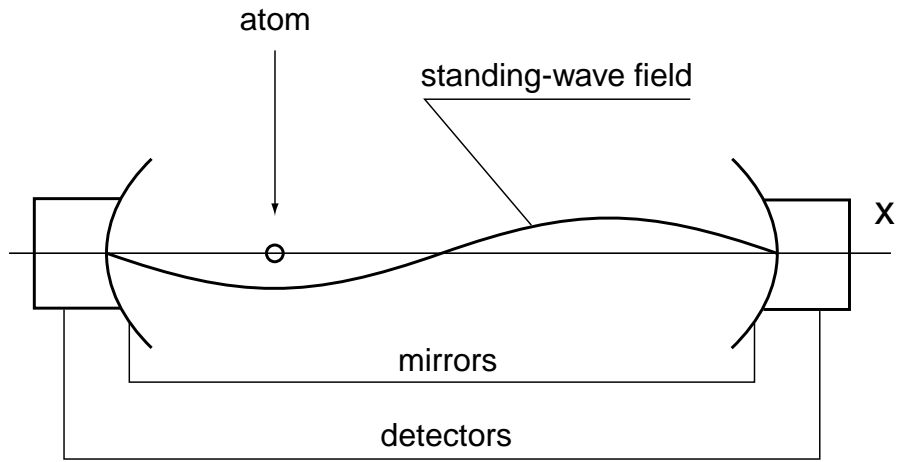}
\end{center}
\caption{The schematic diagram shows a standing-wave microcavity with
detectors.}
\label{fig5}
\end{figure}

In Fig.~\ref{fig5} we depict the scheme of a {\it gedanken} experiment that consists
of a Fabry-Perot optical microcavity with two detectors and cold atoms to be
placed in the cavity. To avoid complications that are not essential to the
main theme of this work, we consider the cavity with only two standing-wave
lengths. Atoms, one by one, are placed at the point $ x_0=0$ with
different values of the detuning $\Delta$. We measure a
time when an atom reaches one of the detectors, the exit time $T$, and
study the dependence $T(\Delta)$ under the other equal conditions
imposed on the atom and the cavity field. 
\begin{figure}
\begin{center}
\includegraphics[width=0.7\textwidth,clip]{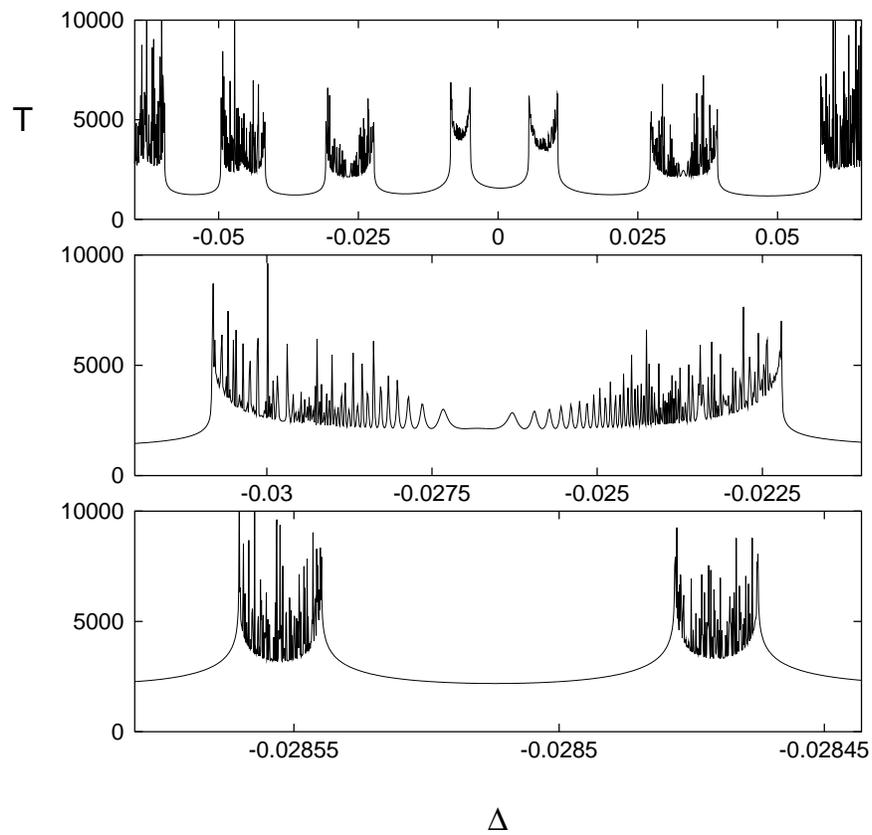}
\end{center}
\caption{Atomic fractals. Exit time of cold atoms $T$ vs the detuning $\Delta$:
$p_0=200$, $z_0=-1$, $u_0=v_0=0$.}
\label{fig6}
\end{figure}
Fig.~\ref{fig6} shows
the function $T(\Delta)$ for atoms with the initial momentum $p_0=200$
($v_a\approx 0.7$ m/s).
The exit time function demonstrates an intermittency of smooth
curves and complicated structures that cannot be resolved in principle, no
matter how large the magnification factor. The middle and low panels in
Fig.~\ref{fig6}
show successive magnifications of the function for the small intervals.
Further magnifications reveals a self-similar fractal-like structure that
is typical for Hamiltonian systems with chaotic scattering \cite{Ott}. 

The exit time
$T$, corresponding to both smooth and unresolved $\Delta$ intervals, increases
with increasing the magnification factor. Theoretically, there exist atoms
never reaching the detectors inspite of the fact that they have no obvious
energy restrictions to leave the cavity. Tiny interplay between chaotic external
and internal dynamics prevents these atoms from leaving the cavity. The similar
phenomenon in chaotic scattering is known as {\it dynamical trapping}. 
In \cite{JETP} for the similar fractal we have computed the
Hausdorff dimension and shown that it is not integer. 

Different kinds of atomic trajectories before detection can be
characterized by the number $m-1$ of changing the sign of momentum.
An $m$-th trajectory corresponds to the atom which
changes the direction of motion before being detected $m-1$ times.
There are also special separatrix-like
trajectories following which atoms asymptotically approach to the points
with the maximum of potential energy, having no more kinetic energy to
overcome it. In difference from the separatrix motion in the resonant system ($\Delta=0$)
with the initial atomic momentum $p_{cr}$, a detuned atom can
asymptotically reach one of the stationary points even after several
oscillations in a well. Let us define the $mS$-trajectory as a trajectory
when the atom changes the direction of motion $m-1$ times and then
begin the separatrix-like motion. Such asymptotical motion
takes the infinite time, so the atom will never be detected.

The smooth $\Delta$ intervals in the first-order
structure (Fig.~\ref{fig6}, upper panel) correspond to atoms which never
changes the direction of motion, i. e. $m=1$, and reaching the right detector.
The unresolved singular
points in the first-order structure with $T=\infty$ at the border between the
smooth and unresolved $\Delta$ intervals are generated by the
$1S$-trajectories. Analogously, the smooth and unresolved $\Delta$ intervals
in the second-order structure (Fig.~\ref{fig6}, middle panel) correspond to 
the 2-nd order and the other trajectories, respectively, with singular points between them
corresponding to the $2S$-trajectories and so on.

There are two different mechanisms of generation of infinite detection times,
namely,
dynamical trapping with infinite oscillations ($m=\infty$) in a cavity and the
separatrix-like motion ($m\ne\infty$). The set of all detunings generating
the separatrix-like trajectories is a countable fractal. Each point in the set
can be specified as a vector in a Hilbert space with $m$ integer nonzero
components. One is able to prescribe to any unresolved interval of
$m$-th order structure a set with $m$ integers, where the first integer is a
number of a second-order structure to which trajectory under consideration
belongs in the first-order structure, the second integer is a number of a
third-order structure in the second-order structure mentioned above, and so on.
Such a number set is analogous to a directory tree address:
"$<$a subdirectory of the root directory$>$/$<$a subdirectory of the 2-nd
level$>$/$<$a subdirectory of the 3-rd level$>$/...". Unlike the separatrix fractal, the set
of all detunungs leading to dynamically trapped atoms with
$m=\infty$ seems to be uncountable. 

\begin{figure}
\begin{center}
\includegraphics[width=0.8\textwidth,clip]{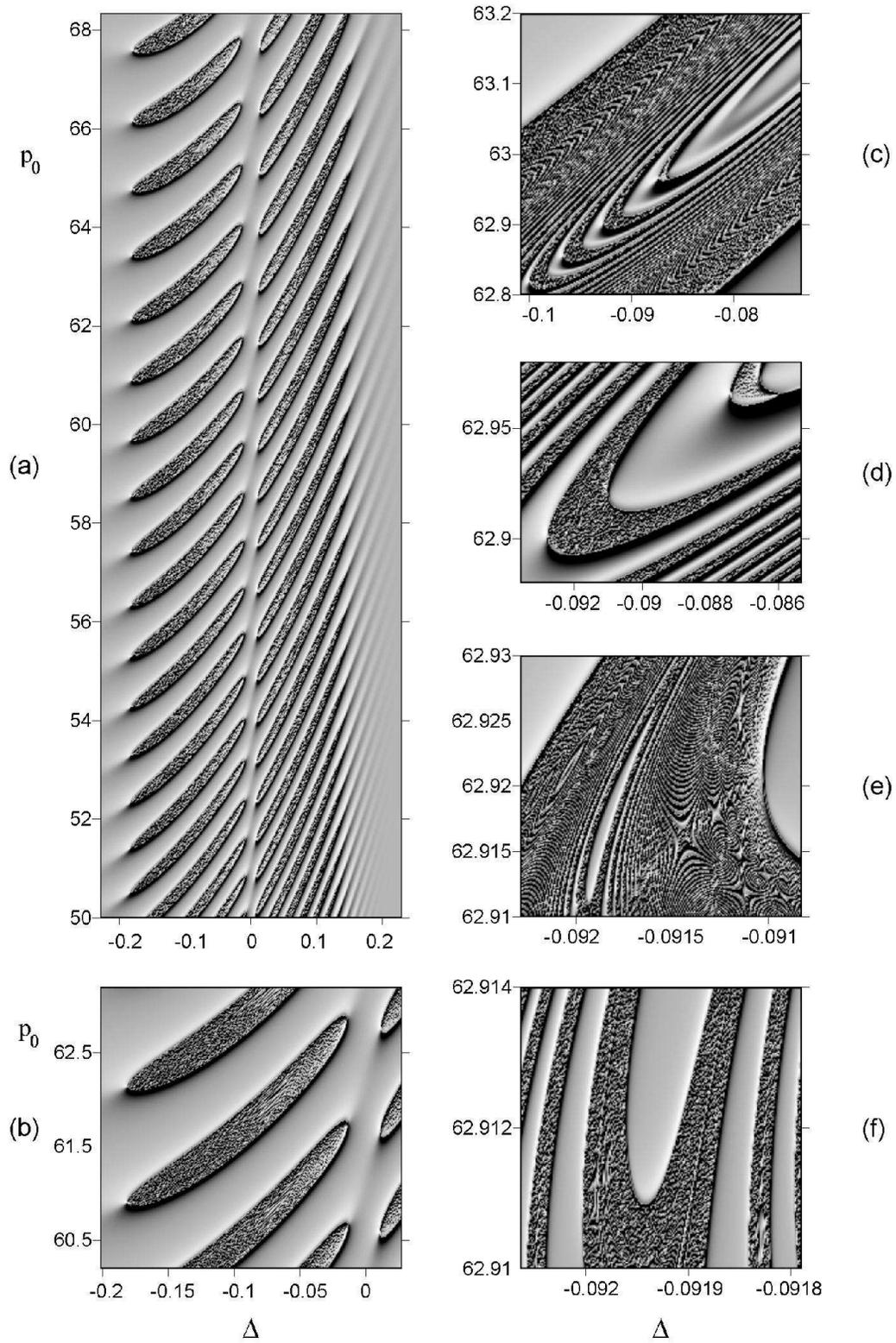}
\caption{Fractal function of the exit time $T$ vs the detuning $\Delta$
and the initial momentum $p_0$. The function is shown in a shaded
relief regime. $\omega_r=9.17\cdot 10^5$.}
\label{fig7}
\end{center}
\end{figure}

The scattering function in the regime of chaotic wandering, time of exit  
$T$, depends  in a complicated way not only on the control parameters but 
initial conditions as well. In Fig.~\ref{fig7} we demonstarte the view 
of this function, whose values are modulated by color, in two coordinates, 
the initial atomic momentum $p_0$ and 
the atom-fireld detuning $\Delta$. From the fragment (a) to the fragment (f) 
we increase subsequently the resolution. One can see increasing complexity 
of the scattreing function with a prominent self-similarity. 
The computation has been performed with the recoil frequency 
$\omega_r=9.17\cdot 10^{-5}$. 

\subsection{Poincar\'e mapping}

The five variables in the equations of motion (\ref{mainsys}) minus the two
integrals of motion (\ref{W}) and (\ref{R}) provide motion in a three-dimensional
space. To visualize the motion we use the idea of mapping trajectories
on two-dimensional planes. Since we have no time-periodic perturbations
in our equations of motion (\ref{mainsys}) we cannot map trajectories  through equal
intervals of time provided by a period of a perturbation. However, the
system has a characteristic space period $2\pi$ imposed by the standing wave.
So we map trajectories on a chosen plane at those time moments when atoms
reach
the positions when $\cos x =1$. We close our phase space along the position
variable with the period $2\pi$. The condition $\cos  x =1$ under fixed
values of the  integrals of motion (\ref{W}) and (\ref{R}) defines a closed
two-dimensional surface in the phase space points of which characterize
unambigiously the system's states. In other words, there is a set of points
on this two-dimensional surface which corresponds to each trajectory with
a given value of the energy $W$. This set can be projected onto a plane of
any system's variables except for the position $ x$. Such a projection is,
generally speaking, two-valued because the two-dimensional surface is closed.
However, one can map trajectories in its ``eastern'' and ``western'' parts
separately.

\begin{figure}
\begin{center}
\includegraphics[width=0.8\textwidth,clip]{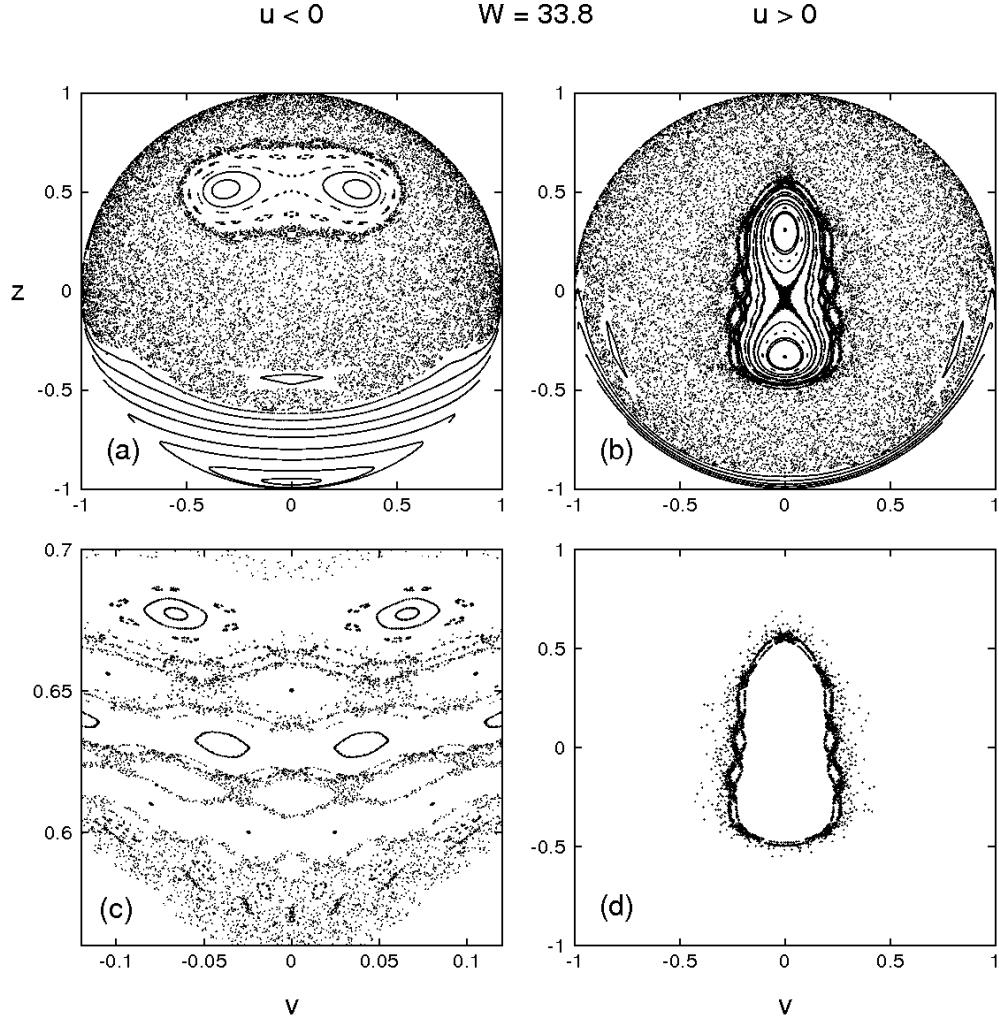}
\end{center}
\caption{Poincar\'e mapping in the Bloch
variable space. (a) $u<0$ (western Bloch hemisphere), (b) $u>0$
(eastern Bloch hemisphere), (c) magnification of the small region
in (a) fragment, (d) mapping with a single chaotic trajectory in 
(b) fragment, illustrating the effect of sticking: $W=33.8$,
$p_{\rm {eff}}=2600$, $\omega_r=10^{-5}$, $\Delta = -0.05$.}
\label{fig8}
\end{figure}

\begin{figure}
\begin{center}
\includegraphics[width=0.8\textwidth,clip]{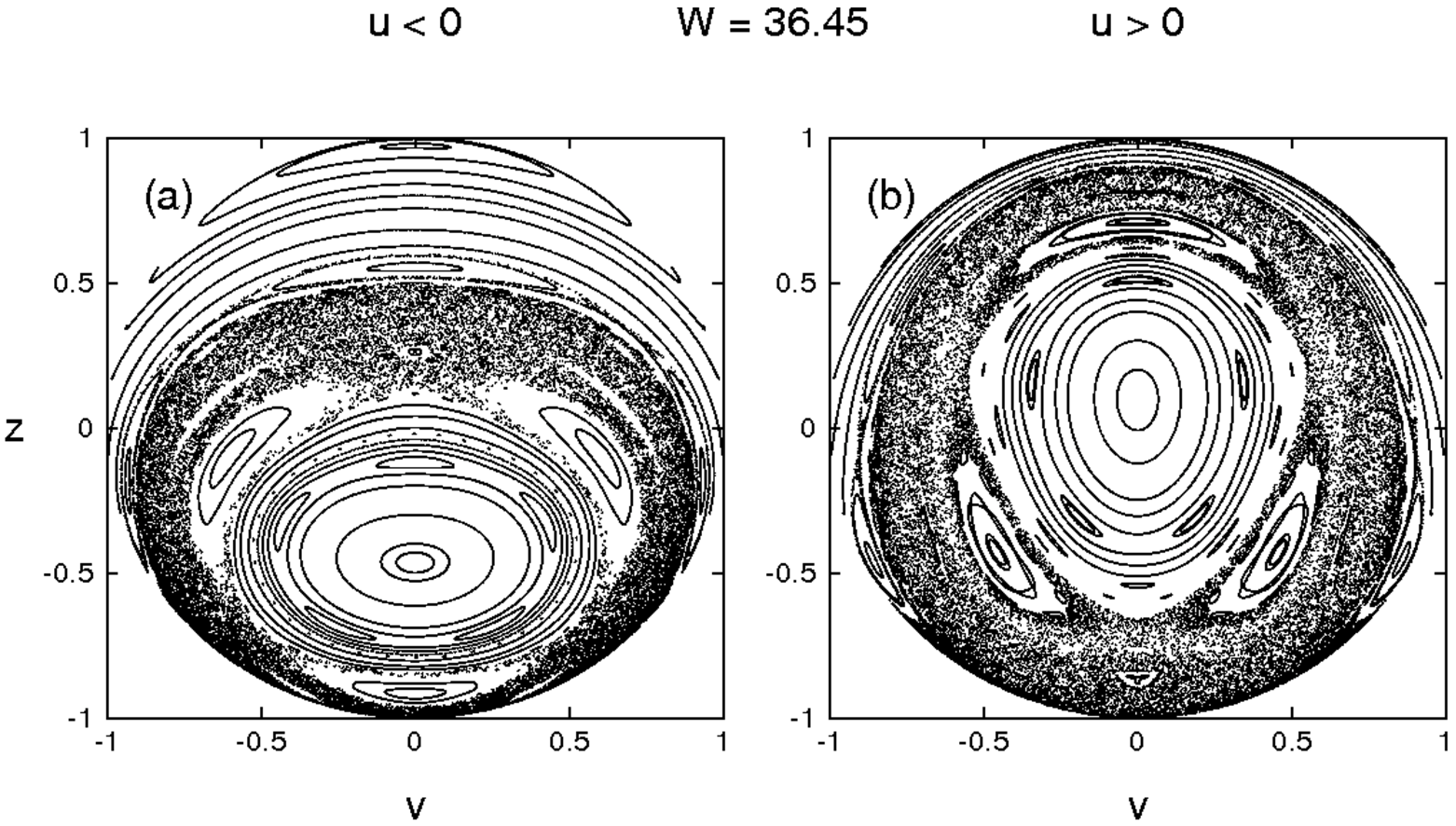}
\end{center}
\caption{Poincar\'e mapping in the Bloch
variable space. (a) $u<0$ (western Bloch hemisphere), (b) $u>0$
(eastern Bloch hemisphere). The parameters are the same as in
Fig.~\ref{fig8}, but $W=36.45$, $p_{\rm {eff}}=2700$.}
\label{fig9}
\end{figure}

In Fig.~\ref{fig8}a and b we demonstrate the Poincar\'e mappings of a number of
atomic trajectories
in the western ($u<0$) and eastern ($u>0$) hemispheres of the Bloch
sphere $(u, v, z)$ on the plane $v-z$, respectively. We fix the values of
the detuning $\Delta = -0.05$,
the recoil frequency $\omega_r = 10^{-5}$, the total energy
$W=33.8$, the initial position $x_0=0$, and map the trajectories with
different other initial conditions, which are restricted by (\ref{W}) and (\ref{R}).
All the mappings were obtained with ballistic atoms whose momenta
slightly (but chaotically for some initial conditions) oscillate around the effective value
$p_{\rm {eff}}=2600$ that corresponds to the chosen value of the energy.
It is such a value of the  momentum which an atom has at the moments when its
potential energy $U$ is zero. 
In Fig.~\ref{fig9} we demonstrate the Poincar\'e mappings of 
a number of atomic trajectories 
in the western ($u<0$) and eastern ($u>0$) hemispheres of the Bloch 
sphere $(u,v,z)$ on the plane $v-z$ just like as in Fig.~\ref{fig8} with 
$W=33.8$ but 
with another value of the total energy $W=36.45$ and the effective 
momentum $p_{\rm {eff}}=2700$. A series of  
bifurcation occurs just between these values of energy and we get in the end 
a central critical point instead of a saddle. One can see 
a typical structure with surviving nonlinear resonances of different orders 
around the center point and overlapping resonances.

In Figs.~\ref{fig8} and \ref{fig9}  
a general views of the mappings in the western and eastern
hemispheres are shown. The pictures are rather typical with chaotic
Hamiltonian systems \cite{Zas}. We see regions of regular motion in the form
of islands and chains of islands filled by regular trajectories which are
known
as Kolmogorov-Arnold-Moser (KAM) invariant curves. The islands are imbedded
into a stochastic sea, and they are produced by nonlinear resonances of
different orders. Increasing the resolution of the mapping, one can see
that big islands are surrounded by islands of a smaller size each of
which, in
turn, is surrounded by a chain of even more smaller islands, and so on to
infinity.
Stochastic layers of the $\infty$-like form are situated between the islands.
From the physical point of view, they are formed by broken and overlapping
nonlinear resonances. From the mathematical point of view, a stochastic layer
is a heteroclinic structure formed by transversal intersections of stable and
unstable manifolds of hyperbolic stationary points. A fractal-like structure
of generations of islands, a trademark of Hamiltonian chaos, is clearly seen
on projections of motion in both the western and eastern hemispheres. To illustrate
what
happens under increasing the resolution of the Poincar\'e mapping,
we plot in Fig.~\ref{fig8}c a zoom of a small region in the stochastic layer in
Fig.~\ref{fig8}a.

We would like to pay attention to another typical phenomena in
Hamiltonian systems, so called sticking \cite{Zas,K83,CS84,M92,BZ93}.
In Fig.~\ref{fig8}d we demonstrate the phenomenon of sticking in the eastern
Bloch hemisphere. The trajectory shown demonstrates an intermittent type
of motion.
It wanders for a while in the stochastic sea as a chaotic trajectory, whose
instability is characterized by a positive value of the finite-time maximal
Lyapunov exponent. Then it is sticked to the boundaries of the outmost
visible
chain of regular islands, where it is practically regular with zero value of
the respective finite-time maximal Lyapunov exponent. It may take a large
amount
of time to find a gap in a cantori structure, surrounding the outmost KAM
tori,
and to get off in the stochastic sea. The process is repeated as time grows.
It should be stressed that sticking influences strongly transport properties
in Hamiltonian systems giving rise to anomalous diffusion, algebraic tails in
distributions of the Poincar\'e recurrence times and of times and lengths
of atomic flights. 

\begin{figure}
\begin{center}
\includegraphics[width=0.8\textwidth,clip]{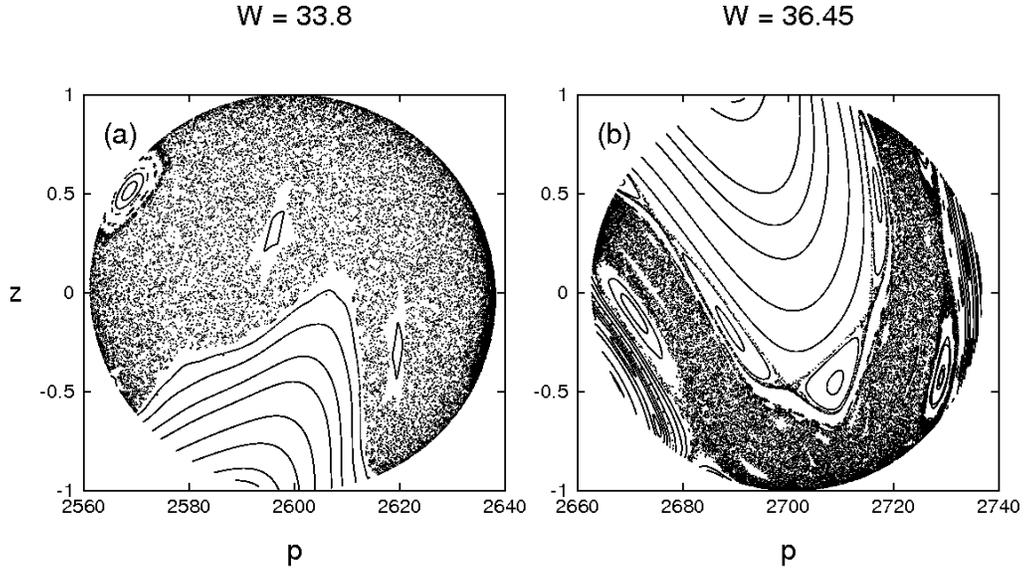}
\end{center}
\caption{Poincar\'e mapping in the
space of the momentum $p$ and the internal energy $z$.
(a) $W=33.8$, $p_{\rm {eff}}=2600$, (b) $W=36.45$, $p_{\rm {eff}}=2700$.
The other values are the same as in Fig.~\ref{fig8}.}
\label{fig10}
\end{figure}
In Fig.~\ref{fig10}a and b we map the same atomic trajectories as in
Figs.~\ref{fig8} and \ref{fig9} onto the plane $p-z$. In this case both the parts
of the closed two-dimensional surface have the same projections because the set
is symmetric with respect to the hyperplane $v=0$. 

\begin{figure}
\begin{center}
\includegraphics[width=0.8\textwidth,clip]{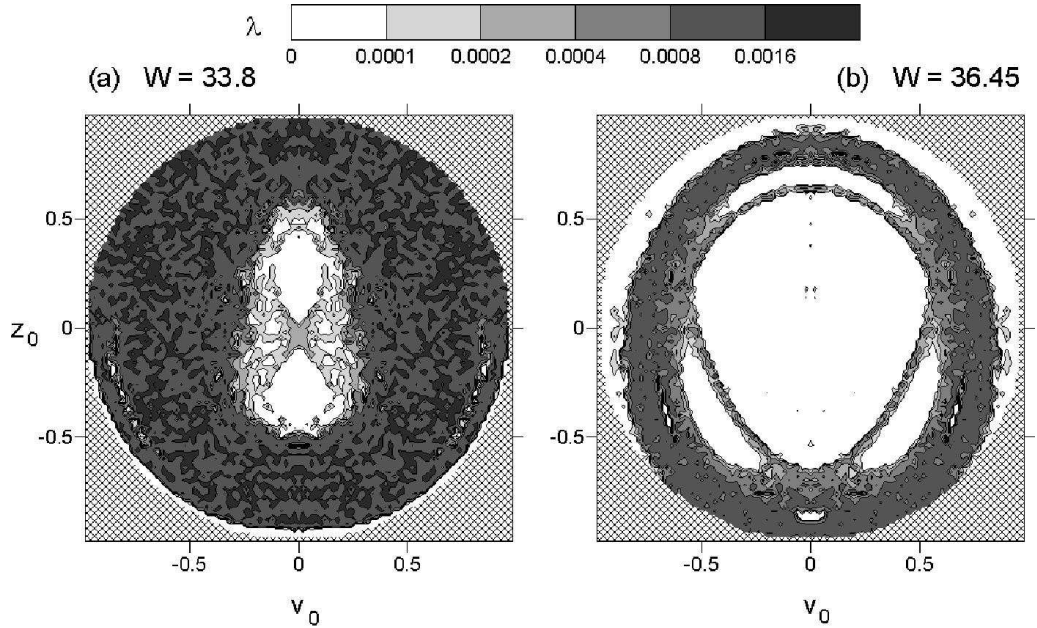}
\end{center}
\caption{The maximal Lyapunov exponent vs the initial values
of the Bloch variables $v_0$ and $z_0$ with $u_0>0$.
(a) $W=33.8$, $p_{\rm {eff}}=2600$, (b) $W=36.45$, $p_{\rm {eff}}=2700$.
The other values are the same as in Fig.~\ref{fig8}.}
\label{fig11}
\end{figure}

In order to quantify instability of the trajectories on 
the Poincar\'e mappings in Figs.~\ref{fig8}  and ~\ref{fig9}, 
we have computed the maps of the maximal Lyapunov exponents exactly with 
the same initial conditions and parameters as in those figures. The results 
in the $v_0-z_0$ coordinates for eastern hemispheres at $W=33.8$  and $W=36.45$ with $u>0$ are shown in 
Figs.~\ref{fig11}a and b, respectively. A rather good correspondence between 
the Poincar\'e mapping and the maximal Lyapunov exponents proves that 
the Poincar\'e mapping we have constructed is a good means to visualize 
compicated dynamics of the coupled internal and external atomic degrees of 
freedom. 

The Poincar\'e mapping with a rich structure of regular and chaotic regions
is typical only in a range of values of the total energy $W$. At the values
$W\gtrsim 40$ atoms move regularly and the respective Poincar\'e mapping
consists of
regular invariant curves only. With decreasing the energy, a series of
bifurcations occurs with appearing of resonant islands of different order.
When decreasing the
energy even more, global chaoticity takes place. At exact resonance
$\Delta =0$,
the value $W=u_0$ corresponds to a separatrix in mechanical variables. Out off
resonance this separatrix is broken, and atoms may wander chaotically in the
optical lattice with a respective irregular Poincar\'e mapping. At
$W\lesssim u_0+\Delta z_0/2$ (including negative values)
atoms are trapped in wells of the optical potential and oscillate there.

\section{Conclusion}

We have considered a simple model of lossless interaction between
a two-level single atom and a  standing-wave single-mode laser field
which creates a one-dimensional optical lattice. Analytical solutions of the
Hamilton-Schr\"odinger equations of motion  have been derived and analyzed
in some limiting cases of regular atomic motion.  Correlations between
the Rabi oscillations  and the center-of-mass motion have been established
and demonstrated. In the regime of chaotic wandering
the atomic motion has been shown to have fractal properties.
Using a special type of the Poincar\'e mapping of atomic
trajectories in an effective three-dimensional phase space onto planes
of atomic internal variables and momentum, we have found
typical structures in Hamiltonian chaotic systems~--- chains of resonant
islands
of different sizes imbedded in a stochastic sea, stochastic layers,
bifurcations,
and so on. The phenomenon of sticking of atomic trajectories to boundaries
of regular islands found in numerical experiments should have a great
influence to atomic transport in optical lattices.

One of the aims of this paper was to describe analytically and numerically
fundamental aspects of nonlinear dynamics of the atom-field interaction.
We have done that  to some extent at the cost of simplifying the model.
To be more realistic we should take into account spontaneous emission events.
In this case, however, the equations of motion cease to be a deterministic
dynamical system because they would include random terms. Our previous
results on  Monte Carlo modelling Hamilton-Schr\"odinger equations have shown
much more complicated type of atomic motion which, besides of chaotic motion,
caused by the fundamental atom-field interaction, includes a purely
stochastic
component caused by random events of spontaneous emission. We plan to  study
the effects of spontaneous emission on chaotic atomic motion
in the future.

\section{Acknowledgments}
This work was supported  by the Russian Foundation for Basic Research
(project no. 06-02-16421 ``Quantum nonlinear dynamics of cold atoms in
        an optical lattice''), by the Program "Mathematical methods in
nonlinear dynamics"
of the Prezidium of the Russian Academy of Sciences (the project
``Dynamical chaos
and coherent structures''), and the program of the Prezidium
of the Far-Eastern Division of the Russian Academy of Sciences
(the projects ``Nonlinear quantum electrodynamics of atoms and photons'' and
``Dissipative dynamics of cold atoms in optical lattices'').

\end{document}